\documentclass{aa}  

\usepackage{graphicx}
\usepackage{txfonts}
\usepackage{hyperref}

\begin{document} 

\title{Time-distance helioseismology: A new averaging scheme for measuring flow vorticity}

\titlerunning{Flow vorticity with time-distance helioseismology}
\authorrunning{}

\author{J. Langfellner \inst{1}
\and
L. Gizon \inst{2,1}
\and
A.~C. Birch \inst{2}}

\institute{Georg-August-Universit\"at, Institut f\"ur Astrophysik, Friedrich-Hund-Platz 1, 37077 G\"ottingen, Germany
     \and
             Max-Planck-Institut f\"ur Sonnensystemforschung,
             Justus-von-Liebig-Weg 3, 37077 G\"ottingen, Germany
             }

\date{Received <date> / Accepted <date>}

\abstract
 {Time-distance helioseismology provides information about vector flows in the near-surface layers of the Sun by measuring wave travel times between points on the solar surface. Specific spatial averages of travel times have been proposed for distinguishing between flows in the east-west and north-south directions and measuring the horizontal divergence of the flows. No specific measurement technique has, however, been developed to measure flow vorticity.}
 {Here we propose a new measurement technique tailored to measuring the vertical component of vorticity.  Fluid vorticity is a fundamental property of solar convection zone dynamics and of rotating turbulent convection in particular.}
 {The method consists of measuring the travel time of waves along a closed contour on the solar surface in order to approximate the circulation of the flow along this contour. Vertical vorticity is related to the difference between clockwise and anti-clockwise travel times.}
 {We applied the method to characterize the vortical motions of solar convection using helioseismic data from the Helioseismic and Magnetic Imager onboard the Solar Dynamics Observatory (SDO/HMI) and from the Michelson Doppler Imager onboard the Solar and Heliospheric Observatory (SOHO/MDI).
 Away from the equator, a clear correlation between vertical vorticity and horizontal divergence is detected. Horizontal outflows are associated with negative vorticity in the northern hemisphere and positive vorticity in the southern hemisphere. The signal is much stronger for HMI than for MDI observations. We characterize the spatial power spectrum of the signal by comparison with a noise model. Vertical vorticity at horizontal wavenumbers below $250/R_\odot$ can be probed with this helioseismic technique.}
 {}
   

   \maketitle
   
\section{Introduction}
The Sun exhibits complex flow patterns in the convection zone, such as turbulent convection, differential rotation, and meridional circulation. These flows are important ingredients for understanding global solar dynamics and the dynamo responsible for the solar 22-year magnetic cycle \citep[cf.][]{toomre_2002}.
Fluid vorticity is a fundamental characteristic of fluid dynamics. The interplay between turbulent convection and rotation can generate cyclonic motions with a net kinetic helicity that depends on solar latitude \citep{duvall_2000}. These motions may convert the toroidal magnetic field into a poloidal field \citep{parker}.
Vortices are not confined to convective motions. \citet{hindman_2009} detected that inflows into active regions \citep{gizon_2001} have a cyclonic component that is presumably caused by solar rotation.

\citet{duvall_1993} showed that near-surface solar flows can be measured using time-distance helioseismology. The idea is to measure the time it takes for solar waves to travel between two surface locations from the temporal cross-covariance of the observable measured at these locations. Typically, the observable is a series of line-of-sight Doppler velocity images, $\phi(\vec{r},t)$, which has been filtered in the Fourier domain to select particular wave packets.
We consider a pair of points $\vec{r}_1$ and $\vec{r}_2$ (``point-to-point geometry''). The cross-covariance $C$ at time lag $t$ is
\begin{equation}
 C(\vec{r}_1,\vec{r}_2,t) = \frac{h_t}{T} \sum_{i=-N}^N \phi(\vec{r}_1,t_i) \phi(\vec{r}_2,t_i+t) ,
\label{eq_cov}
\end{equation}
where $h_t$ is the temporal cadence, $T = (2N+1) h_t$ the observation time, and $t_i = i h_t$ with $i = -N,-N+1,\dots,N$ are the times at which the observable is sampled. From the cross-covariance, the travel time can be measured by fitting a wavelet \citep[e.\,g.,][]{duvall_1997} or a sliding reference cross-covariance \citep{gizon_2004}.

Waves are advected by the flow field $\vec{v}(\vec{r})$, and travel times are sensitive to flows in the vicinity of the ray connecting the points $\vec{r}_1$ and $\vec{r}_2$.
If the flow has a component in the direction $\vec{r}_2 - \vec{r}_1$, then the travel time from $\vec{r}_1$ to $\vec{r}_2$ (denoted by $\tau^+(\vec{r}_1,\vec{r}_2)$) is reduced, while the travel time from $\vec{r}_2$ to $\vec{r}_1$ (denoted by $\tau^-(\vec{r}_1,\vec{r}_2)$) is increased.

\begin{figure}
\centering
\includegraphics[scale=0.2]{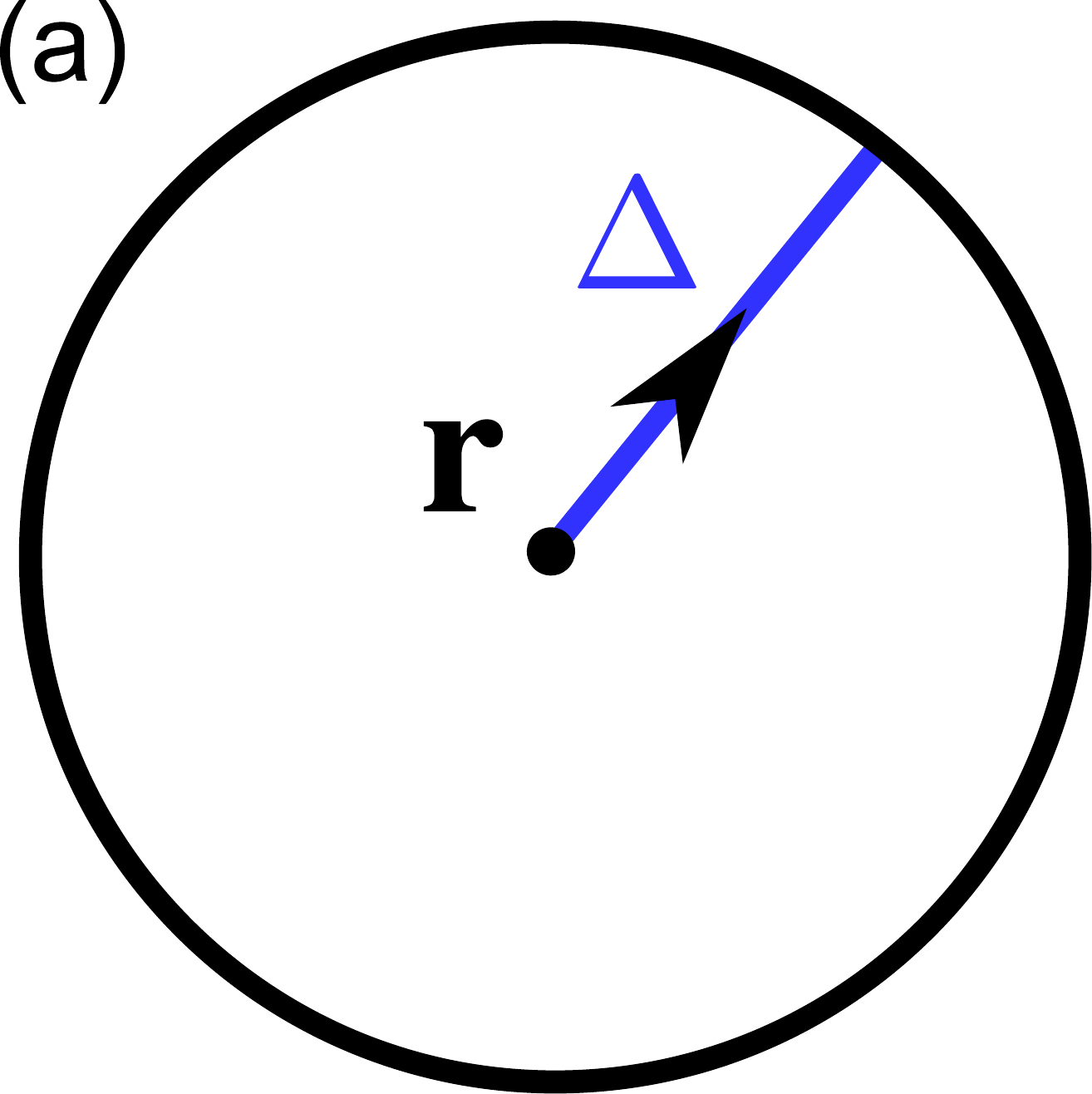} \\ \vspace{0.5cm}
\includegraphics[scale=0.2]{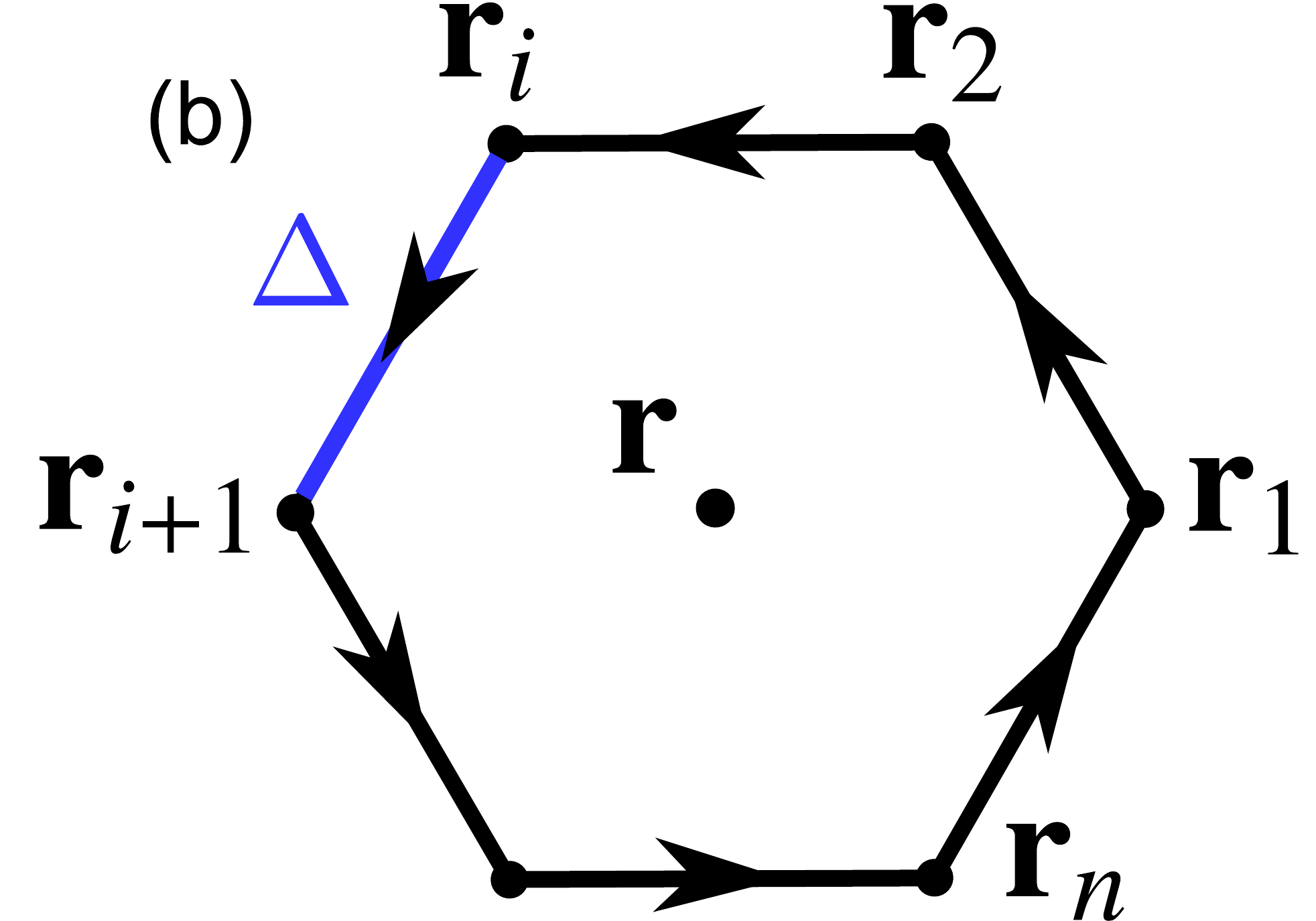} \hspace{1cm}
\includegraphics[scale=0.2]{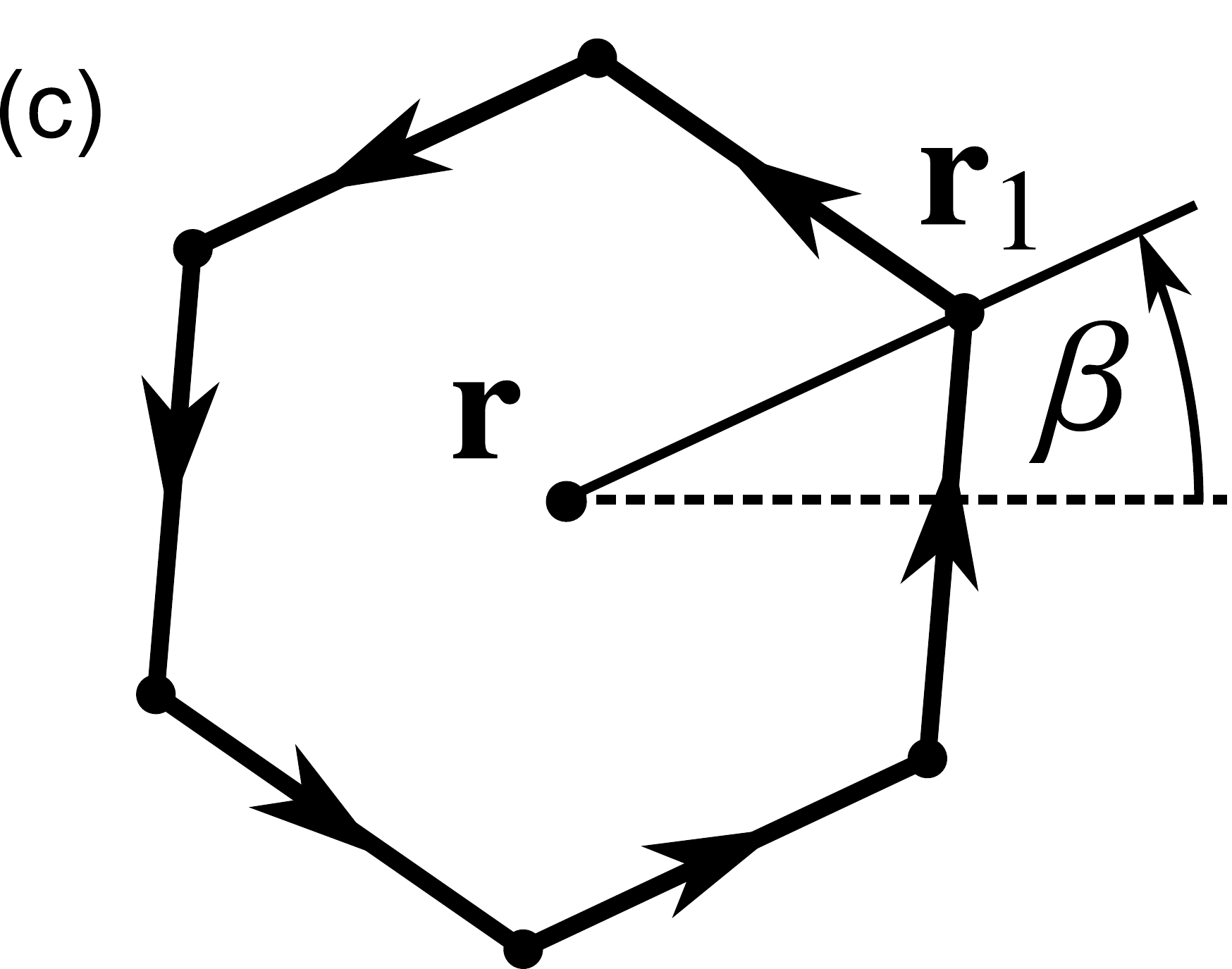}
  \caption{Travel-time measurement geometries. (a) Measurement geometry sensitive to the horizontal component of the flow divergence. Travel times are measured between a central point $\vec{r}$ and the average over a surrounding annulus with radius $\Delta$ as introduced by \citet{duvall_1993}. (b) Proposed new measurement geometry sensitive to the vertical component of flow vorticity. Travel times are measured sequentially along neighboring pairs of points $\vec{r}_i$ and $\vec{r}_{i+1}$ located on a closed contour. In this example, $n=6$ points are used, forming a regular hexagon. (c) As (b), but the hexagon is rotated by an angle $\beta$ around $\vec{r}$.}
\label{fig_1}
\end{figure}

To obtain a measurement that is particularly sensitive to the horizontal flow divergence $\text{div}_h$, travel times are measured between a central point $\vec{r}$ and a surrounding annulus with radius $\Delta$ \citep{duvall_1993}. This ``point-to-annulus geometry'' is displayed in Fig.~\ref{fig_1}a. The flow divergence is related to the difference between inward and outward travel times.

\citet{duvall_1997} proposed to break the annulus into four quadrants pointing in the east, west, north, and south directions, respectively. Here we remind the reader that the solar convention is that west is in the prograde direction of solar rotation. The travel time measured between $\vec{r}$ and the west (or the east) quadrant (``point-to-quadrant geometry'') is sensitive to the component of the flow velocity in the west direction, $v_x$. 
In practice, the difference of the quadrants is used.
In the same fashion, the north component of the flow velocity, $v_y$, can be obtained using the north and south quadrants.

There is no specific measurement geometry, however, that is directly sensitive to the flow vorticity. So far, the vertical component of flow vorticity, $\omega_z = \partial_x v_y - \partial_y v_x$, has been estimated by taking spatial derivatives of the west-east and north-south travel times \citep[see, e.\,g.,][]{gizon_2000}. Alternatively, one could take the spatial derivatives of inverted flow velocities. We would like though a travel-time measurement that is close to the vorticity before performing any inversion. Furthermore, taking derivatives of noisy quantities (as in both cases above) is a dangerous operation. Thus it is desirable to have a travel-time measurement geometry that is explicitly tailored to measure vorticity and that avoids numerical derivatives.

\section{Measuring vortical flows along a closed contour}
In this paper, we implement a measurement technique that is used in ocean acoustic tomography \citep{munk}, where wave travel times are measured along a closed contour $\mathcal C$. This measurement returns the flow circulation along the contour. The flow circulation is related to the vertical component of vorticity $\omega_z$ (averaged over the area $A$ enclosed by the contour) by Stokes' theorem:
\begin{equation}
 \langle \omega_z\rangle_A = \frac{1}{A}\int_A (\vec{\nabla} \times \vec{v}) \cdot \text{d}\vec{A} = \frac{1}{A}\ointctrclockwise_{\mathcal{C}} \vec{v} \cdot \text{d}\vec{l} \ ,
\label{eq_stokes}
\end{equation}
where $\vec{v}$ is the flow velocity vector on the surface $A$. The vector $\text{d}\vec{A}$ is normal to the solar surface (upward) and the contour integral runs anti-clockwise with $\text{d}\vec{l}$ tangential to the contour.

\subsection{Geometry for anti-clockwise travel times}
We approximate the contour integral as follows. We select $n$ points $\vec{r}_1, \vec{r}_2,\dots,\vec{r}_n$ along a circular contour $\mathcal{C}$ and measure the travel times $\tau^+$ pairwise in the anti-clockwise direction. Neighboring points are each separated by an equal distance $\Delta$. The points form the vertices of a regular polygon (Fig.~\ref{fig_1}b). Averaging over the $\tau^+$ measurements yields what we call the ``anti-clockwise travel time'' $\tau^\circlearrowleft$,
\begin{equation}
 \tau^\circlearrowleft(\vec{r},\Delta,n) := \frac{1}{n} \sum_{i=1}^n \tau^+(\vec{r}_i,\vec{r}_{i+1}) \ ,
 \label{eq_tauccw-def}
\end{equation}
with the notation $\vec{r}_{n+1} = \vec{r}_1$.
With this definition, $\tau^\circlearrowleft$ is reduced when there is a flow velocity $v_\phi$ tangential to the circle of radius $R=\Delta/[2\sin(\pi/n)]$ in anti-clockwise direction. In order to provide a simplified description of the relationship between $\tau^\circlearrowleft$ and $\omega_z$, we may write the perturbation to the anti-clockwise travel time caused by $v_\phi$ as 
\begin{equation}
 \delta\tau^\circlearrowleft \approx -\tau_0 \frac{v_\phi}{V_\text{ref}} \sim -\frac{\tau_0 R}{2V_\text{ref}} \omega_z ,    \label{eq_tau-perturbation}
\end{equation}
where $\tau_0$ is the unperturbed travel time and $V_\text{ref}$ is the reference wave speed. This description only provides a rule-of-thumb connection between $\tau^\circlearrowleft$ and $\omega_z$. The proper relationship is described by 3D sensitivity kernels \citep{birch_2007}.
Additionally, these quantities are functions of all three spatial dimensions, which has not been accounted for.

We note that the distance $\Delta$ between the points must be greater than the wavelength (e.\,g., about $5$~Mm for the f mode at 3~mHz), in order to distinguish waves propagating from $\vec{r}_i$ to $\vec{r}_{i+1}$ from waves propagating in the opposite direction.
Also there is some freedom in selecting the contour $C$. For example, active regions are often shaped irregularly. To measure the vorticity around active regions, it may be useful to adapt the contour to the shape of the active region.

\subsection{Reducing the noise level}
\begin{figure*}
  \centering
\includegraphics[width=\hsize]{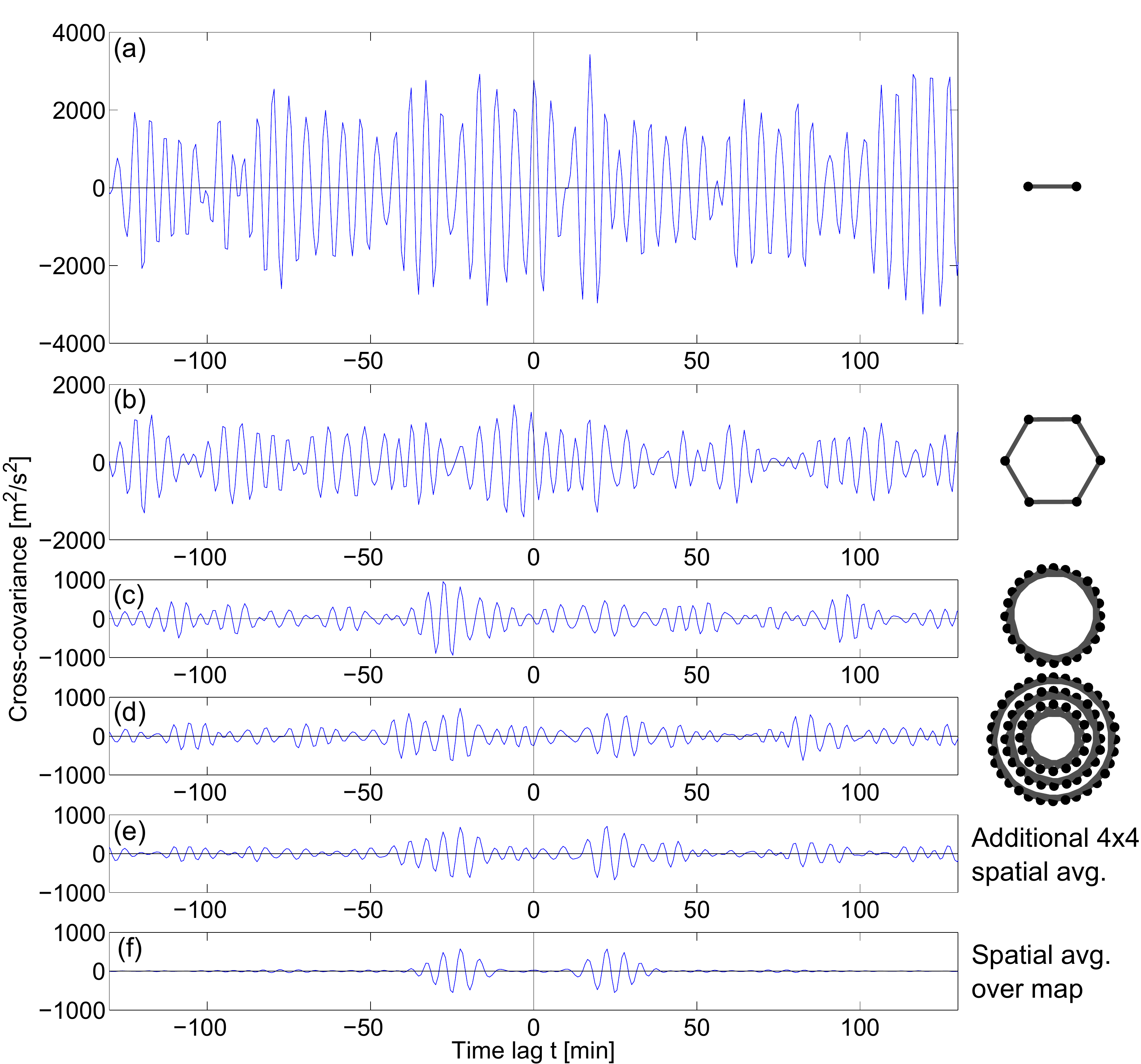}
      \caption{F-mode cross-covariance $C$ as a function of time lag for HMI data (see Sect.~\ref{chap_observations}). All curves except (e) and (f) are for a single center point $\vec{r}$ near disk center (8~h averaging, 1 May 2010 dataset). The distance $\Delta$ between consecutive points is 10~Mm. (a) $C$ for a single pair. (b) Averaging over annulus with $n=6$. (c) Further averaging over eight angles $\beta$. (d) Further averaging over two additional radii: $n=4$ with four angles $\beta$ and $n=8$ with eight angles $\beta$. (e) Further averaging over annuli centers ($4\times4$ pixels). (f) For reference, $n=6$ annulus averaged over an entire map (about $180\times180~$Mm$^2$).}
\label{fig_2}
    \end{figure*}

The definition of the anti-clockwise travel times (Eq.~(\ref{eq_tauccw-def})) assumes that pairwise travel times can be measured, irrespective of the noise level. This is not a problem in the quiet Sun using the travel-time definition of \cite{gizon_2004}, which is very robust with respect to noise. However, a wavelet fit to the cross-covariance \citep[as in][]{duvall_1997} is only possible when the noise is sufficiently low. This is not the case for a single pair of points (see Fig.~\ref{fig_2}a for an example using f-mode-filtered SDO/HMI data with $\Delta = 10\,$Mm).

One option is to average $C$ before performing the wavelet fit. At fixed $\Delta$ and $n$, the measurement polygon can be rotated by angles $\beta$ around $\vec{r}$ (Fig.~\ref{fig_1}c). Since plane waves are only weakly correlated for different propagation directions, taking the average over various angles $\beta$ will lower the noise level. In Fig.~\ref{fig_2}c, an average over eight angles $\beta$ is shown for $n=6$. Furthermore, $C$ can be averaged over several annulus radii $R$ (several $n$ at fixed distance $\Delta$). In Fig.~\ref{fig_2}d, the cross-covariance is averaged over three different annuli ($n=4$, 6, and 8) and angles $\beta$. An additional $4\times 4$ averaging over the centers of annuli (Fig.~\ref{fig_2}e) gives a cross-covariance function that has a sufficiently low level of noise to be analyzed by a wavelet fit. Such averaging procedures are often used for measuring outward$-$inward travel times. Any spatial averaging must be properly taken into account when travel-time inversions are performed later.

\subsection{Decoupling from isotropic wave-speed perturbations}
Since the purpose of $\tau^\circlearrowleft$ is to measure the vorticity $\omega_z$, it should ideally not be sensitive to any other solar perturbation. However, a wave-speed perturbation along the contour due to fluctuations in, for instance, temperature or density will also perturb $\tau^\circlearrowleft$. In order to remove travel-time perturbations that are not caused by the flow, we also measure the clockwise travel time
\begin{equation}
 \tau^\circlearrowright(\vec{r},\Delta,n) := \frac{1}{n} \sum_{i=0}^{n-1} \tau^+(\vec{r}_{n-i+1},\vec{r}_{n-i}) \ .
 \label{eq_taucw-def}
\end{equation}
For example, the travel times $\tau^\circlearrowleft$ and $\tau^\circlearrowright$ are affected in the same way by a temperature perturbation. We thus introduce the difference anti-clockwise minus clockwise travel time (denoted with the superscript ``ac''),
    \begin{equation}
    \tau^\text{ac}(\vec{r},\Delta,n) := \tau^\circlearrowleft(\vec{r},\Delta,n) - \tau^\circlearrowright(\vec{r},\Delta,n) \ .
\label{eq_tau-ccw-diff}
    \end{equation}
Note that $\tau^\text{ac} \approx 2 \delta\tau^\circlearrowleft$, where $\delta\tau^\circlearrowleft$ is given by Eq.~(\ref{eq_tau-perturbation}). The travel time $\tau^\text{ac}$ should be largely independent of perturbations other than a vortical flow.
This approach is similar to the one proposed by \citet{duvall_1997} to measure the flow divergence from point-to-annulus travel times.

\section{Proof of concept using SDO/HMI and SOHO/MDI observations}
In order to test if the proposed averaging scheme is able to measure flow vorticity, we have carried out two simple experiments using f modes. The first experiment (Sect.~\ref{chap_test1}) consists of making maps of $\tau^\circlearrowleft$ using SDO/HMI observations \citep{schou_2012}, computing the spatial power spectrum of these maps, and comparing with the predicted power spectrum of pure realization noise. The second experiment (Sect.~\ref{chap_test2}) is to look for a correlation between vertical vorticity and horizontal divergence. The sign and the amplitude of this correlation are expected to scale like the local Coriolis number.

\subsection{Observations} \label{chap_observations}
We used $112\times24$~h series of SDO/HMI line-of-sight velocity images. The Dopplergrams were taken from 1 May to 28 August 2010 when the Sun was relatively quiet. Regions of the size $180\times180$~Mm$^2$ at solar latitudes from $-60\degr$ to $+60\degr$ in steps of $20\degr$ were tracked for one day as they crossed the central meridian. Images were remapped using Postel's projection and tracked at the local surface rotation rate from \citet{snodgrass_1984}. The resulting data cubes were cut into three 8~h datasets. A ridge filter was applied to select f modes.

We also used $56\times24$~h series of SOHO/MDI \citep{scherrer_1995} full-disk line-of-sight velocity images from 8 May through 11 July 2010, thus overlapping with the HMI observations. The MDI data were processed in the same way as for HMI, however the spatial sampling of MDI is lower by a factor of four (2.0 arcsec px$^{-1}$ instead of 0.5 arcsec px$^{-1}$) and the temporal cadence is 60~s instead of 45~s. The spatial resolution is given by the instrumental point spread function (PSF), which can be approximated by a Gaussian with a full width at half maximum (FWHM) of about $3.3$~arcsec ($2.4$~Mm) for MDI \citep{korzennik_2004,korzennik_2012_erratum} and about $1.0$~arcsec ($0.7$~Mm) for HMI \citep{yeo_2014}.

\subsection{Travel-time maps} \label{chap_tt-maps}
From the f-mode-filtered Dopplergrams, we computed the cross-covariance $C$ in Fourier space,
\begin{equation}
C(\vec{r}_1,\vec{r}_2,\omega) = h_\omega \phi_T^*(\vec{r}_1,\omega) \phi_T(\vec{r}_2,\omega) ,
\end{equation}
where $\omega$ is the angular frequency and $h_\omega = 2\pi/T$ is the frequency resolution. The symbol $\phi_T$ denotes the Fourier transform of $\phi(\vec{r},t)$ multiplied by the temporal window function. This way of computing $C$ is equivalent to Eq.~(\ref{eq_cov}), but is much faster.
To measure the travel times $\tau^\circlearrowleft$ and $\tau^\circlearrowright$, we used the linearized definition of travel times as defined by Eq.~(3) in \citet{gizon_2004} with $W$ given by Eq.~(4) in that paper. We obtained the reference cross-covariance $C^\text{ref}$ by spatially averaging $C$ over the whole map.
We used $\Delta = 10$~Mm and $n = 6$, which corresponds to an annulus radius of $10$~Mm, and used four different values for $\beta$ (0$\degr$, 15$\degr$, 30$\degr$, and 45$\degr$). We computed $\tau^\text{ac}$ as defined in Eq.~(\ref{eq_tau-ccw-diff}). Additionally, we computed outward$-$inward mean travel-time maps ($\tau^\text{oi}$) between an annulus radius of $10$~Mm and the central point. Again, we used the travel-time definition from \citet{gizon_2004}.

\begin{figure*}
  \centering
\includegraphics[width=\hsize]{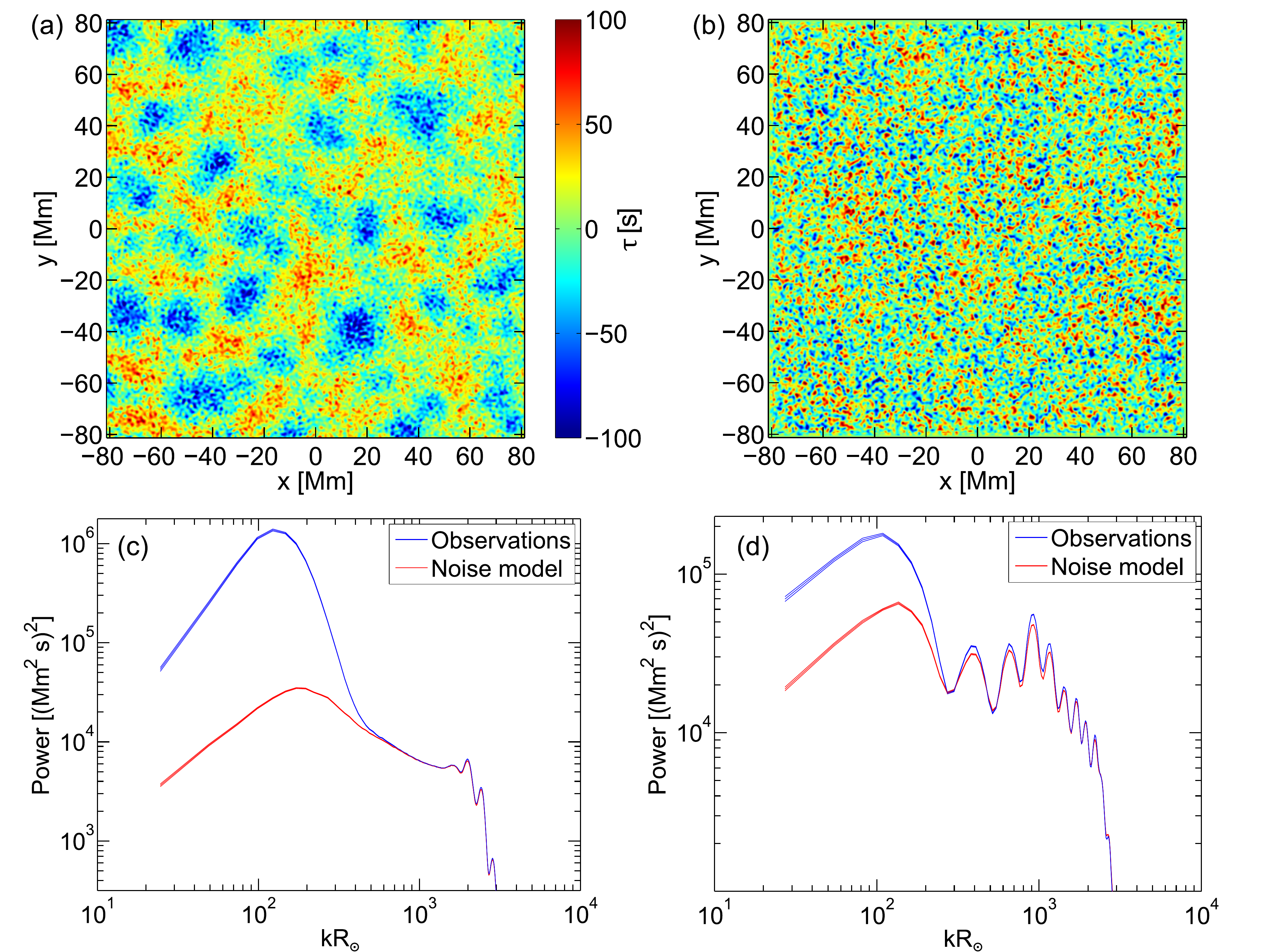}
      \caption{(a) Example HMI $\tau^\text{oi}$ travel-time map for one f-mode-filtered 8~h dataset centered at the solar equator and annulus radius $\Delta = 10~$Mm. The coordinate $x$ is in the west direction and $y$ in the north direction. The travel times have been measured using the linearized definition in \citet{gizon_2004}. The color scale has been truncated to increase the contrast. The minimum and maximum values are $-152.4$~s and $117.3$~s. (b) Example $\tau^\text{ac}$ travel-time map for the same dataset as in (a) with $\Delta = 10~$Mm and $n=6$, averaged over the four angles $\beta$. The minimum and maximum values are $-149.7$~s and $144.8$~s. (c) Power spectrum of $\tau^\text{oi}$ for both HMI data and noise model \citep{gizon_2004}, averaged over azimuth and $336\times8$~h datasets (size about $180\times180$~Mm$^2$) centered at the solar equator, plotted versus the product of horizontal wavenumber $k$ and solar radius $R_\odot$. The thickness of the lines denotes the $1\sigma$ error. (d) Power spectrum of $\tau^\text{ac}$ for both HMI data and noise model, averaged over azimuth and $336\times8$~h datasets. The $\tau^\text{ac}$ maps were averaged over four angles $\beta$ before computing the power.}
\label{fig_3}
    \end{figure*}

Figure~\ref{fig_3}a shows a $\tau^\text{oi}$ travel-time map for an example 8~h dataset at the solar equator. The bluish features of size 20 to 30~Mm are areas of positive divergence. They represent supergranular outflow regions. Conversely, the reddish areas show the supergranular network of converging flows. For the same dataset, a $\tau^\text{ac}$  map that was averaged over the four angles $\beta$ is depicted in Fig.~\ref{fig_3}b. There is no evidence of excess power at the scales of supergranulation.

\subsection{Test 1: Evidence of a vorticity signal in $\tau^\text{ac}$ as a function of wavenumber} \label{chap_test1}
In order to evaluate the signal-to-noise ratio (S/N) in the $\tau^\text{ac}$ maps, we compare the spatial power of the travel-time maps with a noise model. For the noise model, we use the recipe of \citet{gizon_2004} to construct artificial datasets as follows. In 3D Fourier space the observable is modeled by a Gaussian complex random variable with zero mean and variance given by the expected power spectrum (estimated from the observations). In the noise model, wavenumbers and frequencies are uncorrelated to exclude wave scattering by flows and heterogeneities (signal). The expectation value of the power spectrum is chosen to match that of HMI observations. A detailed study and a validation of the noise model is provided by \cite{fournier_2014}.

For each HMI dataset, one realization of the noise model was generated, based on the corresponding power spectrum. From these noise datasets, we computed $\tau^\text{oi}$ and $\tau^\text{ac}$ travel-time maps in the same manner as for the HMI observations (cf.~Sect.~\ref{chap_tt-maps}).

The $\tau^\text{oi}$ power spectrum averaged over azimuth and all datasets is shown in Fig.~\ref{fig_3}c. There is signal above noise level for $kR_\odot < 500$, with a maximum S/N of 50 at $kR_\odot = 120$. This is the well-known supergranulation peak. For high wavenumbers ($kR_\odot > 500$), the travel-time maps are dominated by noise. The contribution from convection features that are much shorter lived than the observation time of 8~h is very small. At $kR_\odot \sim2000$ there is a cut-off in power corresponding to a wavelength of 2.5~Mm (half the f-mode wavelength at 3~mHz).
For MDI (not shown here), the S/N at the supergranulation peak is about 16 and the S/N vanishes at $kR_\odot=400$.

For $\tau^\text{ac}$, we averaged the maps over four angles $\beta$, computed the power spectrum for each resulting map and averaged the power spectra over azimuth and all datasets. The result for HMI is shown in Fig.~\ref{fig_3}d. The power for the HMI travel times significantly exceeds that of the noise model for $kR_\odot < 250$, with a S/N increasing toward larger scales (S/N about 1.5 at $kR_\odot=120$) and reaching a S/N of about 2.6 for $kR_\odot = 25$. This is qualitatively different than for the $\tau^\text{oi}$ power. Again, there is a power cut-off at $kR_\odot \sim2000$.
In the MDI case, the S/N is similar but lower than for the HMI data ($\text{S/N}=0.7$ at $kR_\odot=120$, $\text{S/N}=1.3$ at $kR_\odot=25$).

\begin{figure*}
  \centering
\includegraphics[width=0.49\hsize]{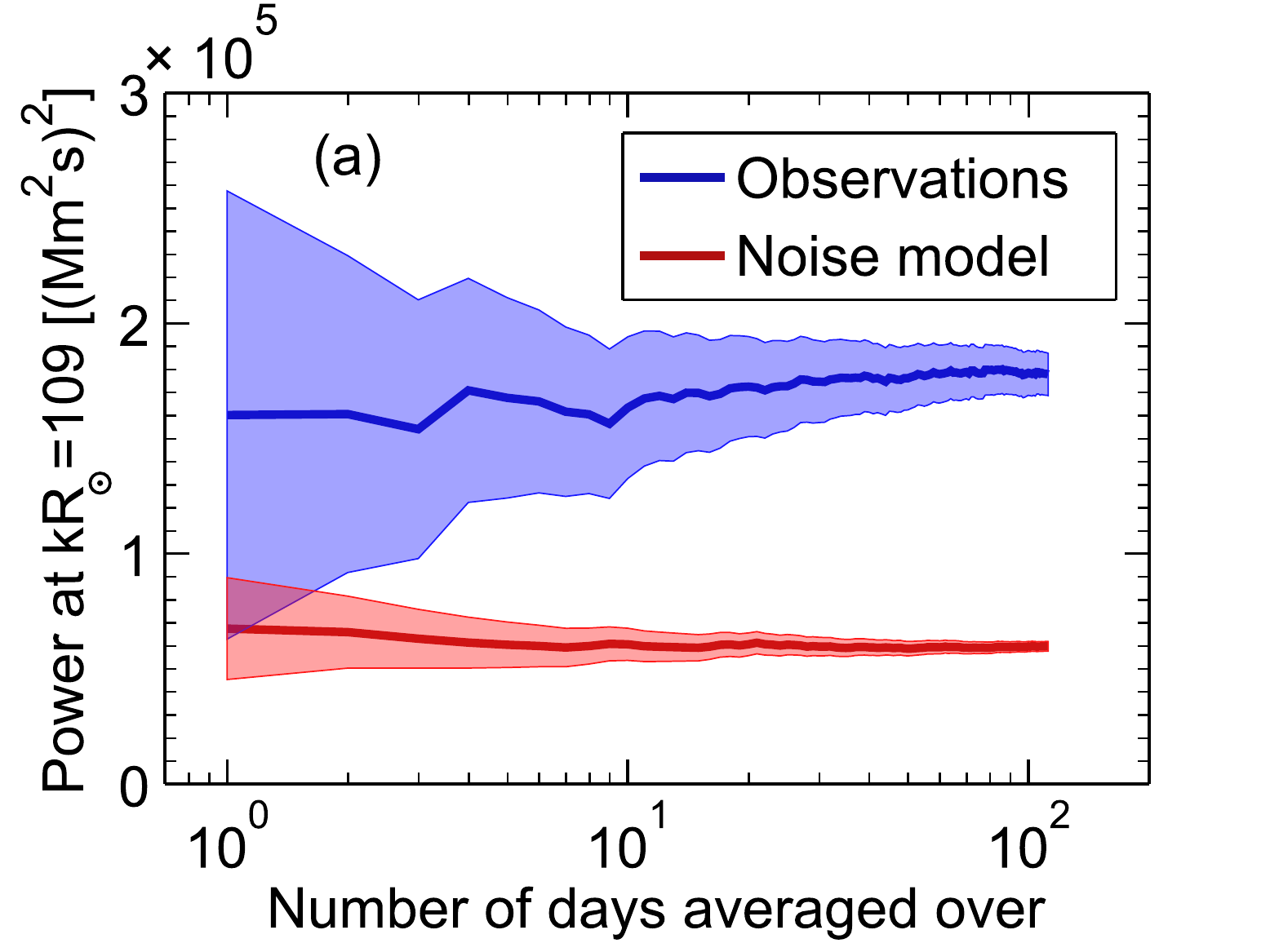}
\includegraphics[width=0.475\hsize]{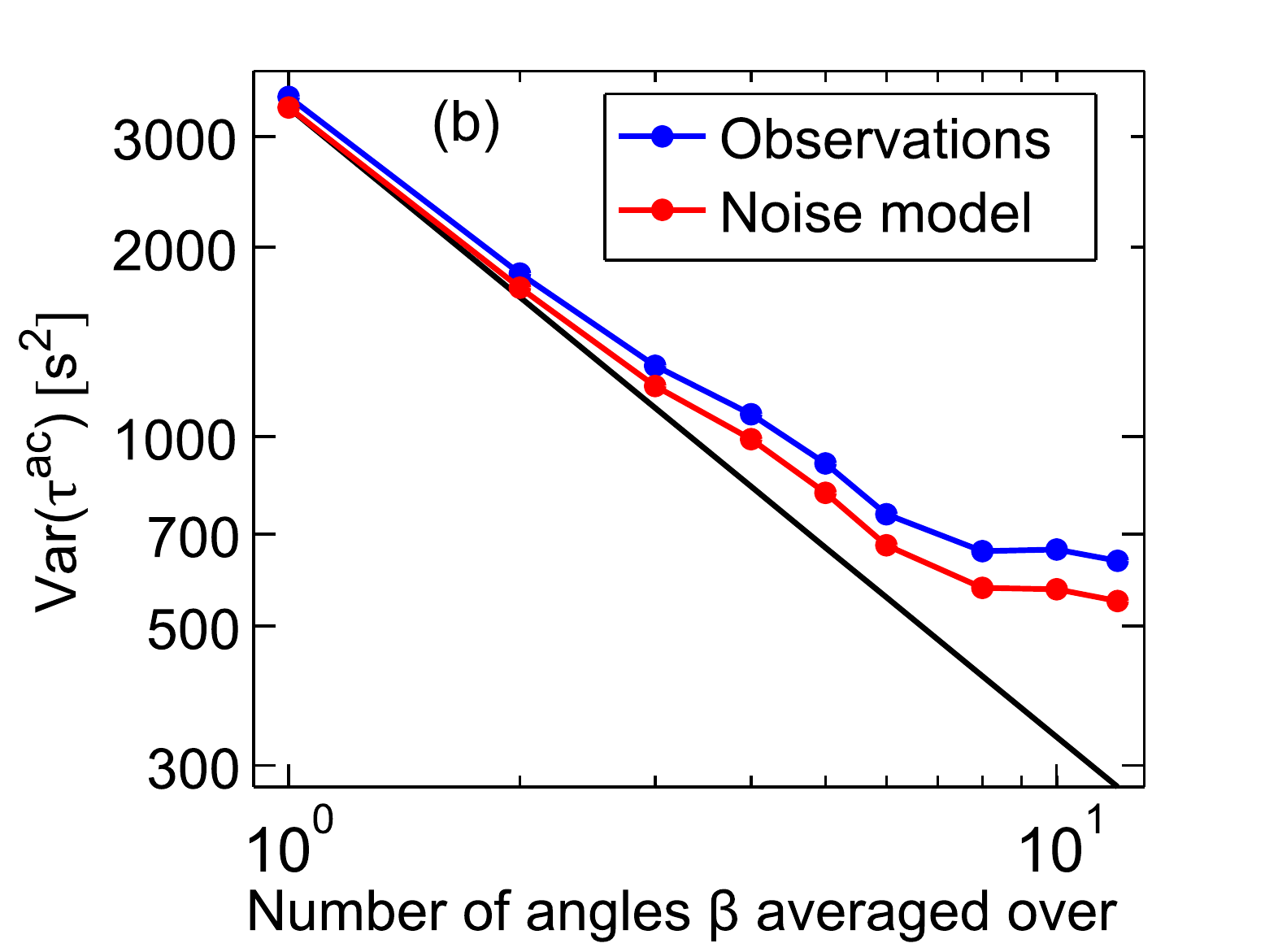}
      \caption{(a) Azimuthally averaged power of HMI and noise-model $\tau^\text{ac}$ maps (average maps over four angles $\beta$) at $kR_\odot = 109$, as a function of days of data averaged over (the same data as in Fig.~\ref{fig_3} were used). The power was averaged over three 8~h datasets per day. The filled areas denote the $3\sigma$ error estimates. (b) Variance of $\tau^\text{ac}$ measurements computed from both HMI and noise-model f-mode-filtered data at the solar equator. The observation time of the datasets is 8~h and their spatial size is about $180 \times 180$~Mm$^2$. The variance is shown as a function of the number of angles $\beta$ over which the data is averaged. For reference, the black line shows $\text{Var}(\tau^\text{ac}) \propto 1/n_\beta$.}
\label{fig_4}
    \end{figure*}

For the results presented in Fig.~\ref{fig_3}d as much as four months of HMI data were averaged. What is the minimum number of days of observations needed to achieve a clear detection of the vorticity signal in the $\tau^\text{ac}$ maps (180 Mm on the side)? By clear detection we mean that, at fixed wavenumber, the power in the observed travel times and the power in the noise model are separated by at least $2 \times 3\sigma$. This requirement is somewhat arbitrary but safe.
Figure~\ref{fig_4}a shows the spatial power of $\tau^\text{ac}$ at $kR_\odot = 109$ as a function of observation duration, for $\tau^\text{ac}$ maps averaged over four angles $\beta$. Overplotted are the $3\sigma$ error estimates (filled areas). Since the data cubes are not correlated from one day to the next (different longitudes), the variance of the noise decreases like 1 over the number of days. The criterion for clear signal detection is fulfilled after two days of observations.

This detection is also a function of the number of angles $\beta$ over which the contours are rotated and averaged. In Fig.~\ref{fig_4}b, the variance of the $\tau^\text{ac}$ travel times is plotted versus this number, $n_\beta$. For $\Delta = 10$~Mm, we find that the variance decreases almost like $1/n_\beta$ for small $n_\beta$ and reaches a plateau for $n_\beta > 8$. This is because the $\tau^\text{ac}$ measurements are highly correlated for small rotation angles. In this case, $n_\beta=4$ is a good compromise between efficient data use and computation time.

\subsection{Test 2: Effect of rotation on vorticity in supergranules} \label{chap_test2}
Here we compute $\langle \tau^\textrm{oi}\tau^\text{ac}\rangle$, where the angle brackets denote an average over the solar surface and over all datasets. Since $\tau^\textrm{oi} \propto -\text{div}_h$ and $\tau^\text{ac} \propto -\omega_z$, the product $\langle \tau^\textrm{oi}\tau^\text{ac}\rangle$ serves as a proxy for $\langle \text{div}_h \omega_z \rangle$, which is a component of the kinetic helicity and is sensitive to the effect of the Coriolis force on convection \citep[e.\,g.,][]{zeldovich,ruediger_1999}.

\begin{figure*}
  \centering
\includegraphics[width=0.46\hsize]{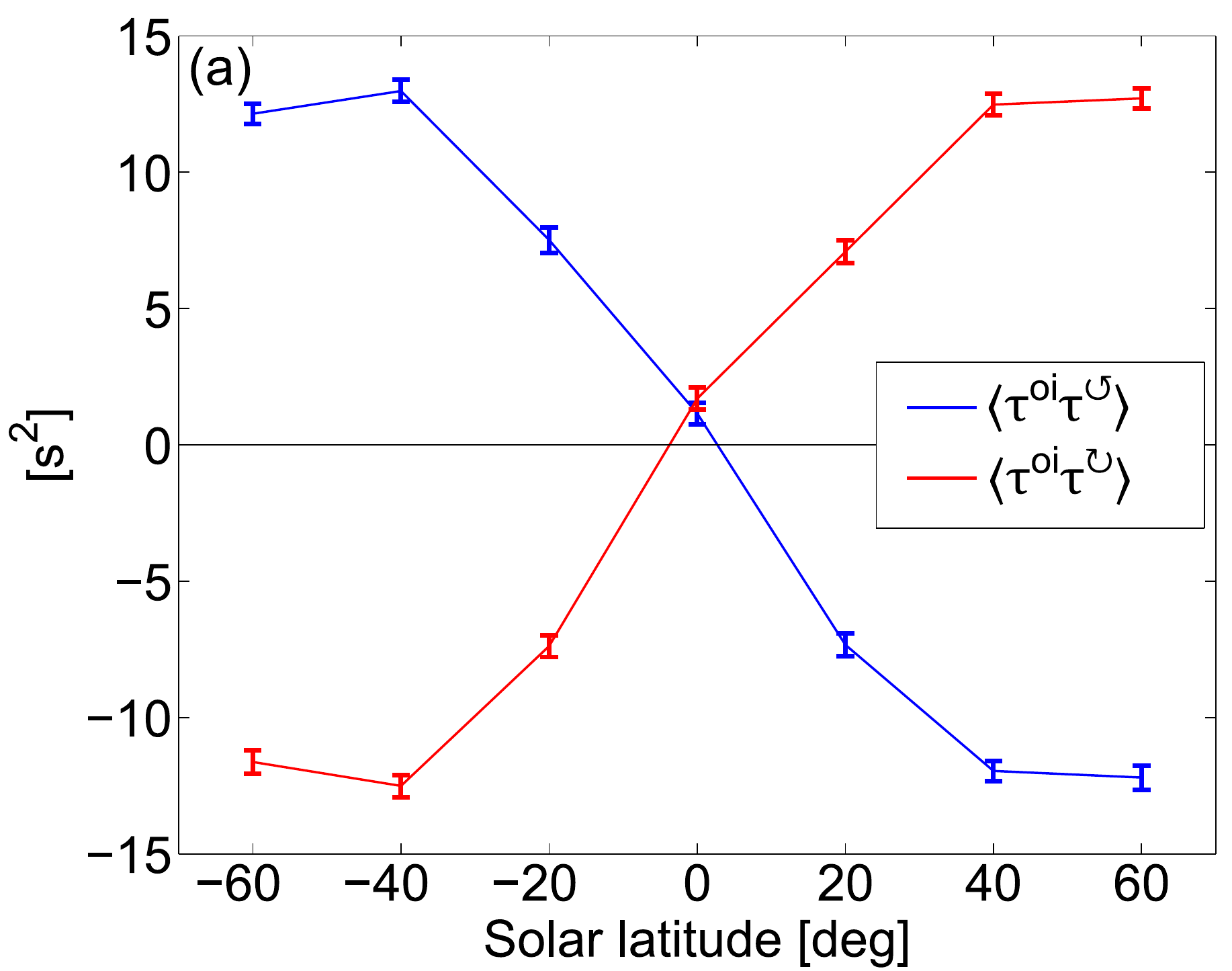}
\includegraphics[width=0.50\hsize]{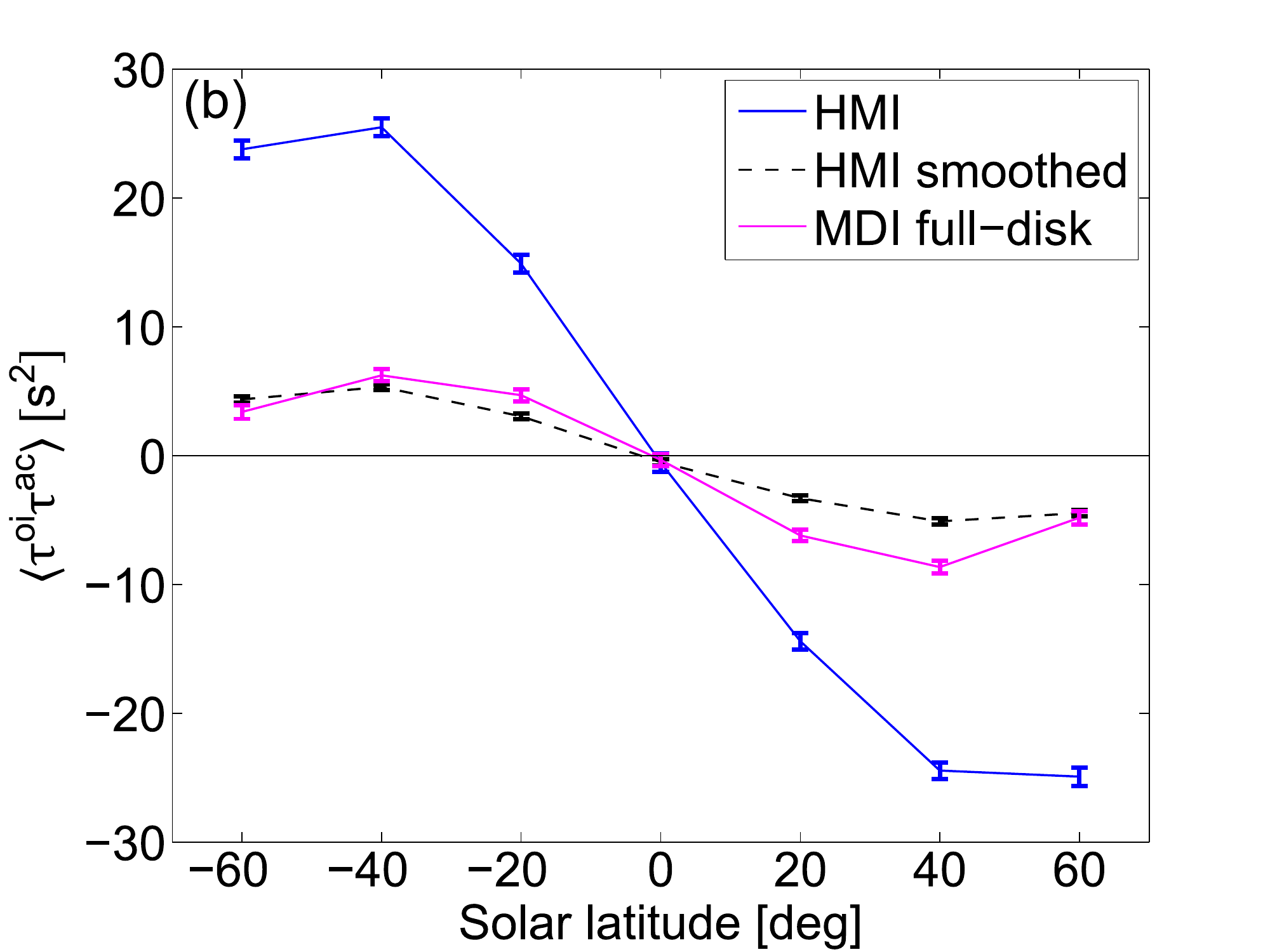}
      \caption{(a) Horizontal averages $\langle \tau^\textrm{oi}\tau^\circlearrowleft\rangle$ and $\langle \tau^\textrm{oi}\tau^\circlearrowright\rangle$ for f-mode-filtered HMI observations as functions of solar latitude, averaged over $336\times8$~h datasets of about $180 \times 180$~Mm$^2$ and four angles $\beta$. (b) Horizontal average $\langle \tau^\textrm{oi}\tau^\text{ac}\rangle$ for different data: HMI at full resolution (0.5 arcsec px$^{-1}$), MDI full-disk data (2.0 arcsec px$^{-1}$), and HMI data spatially averaged over $4 \times 4$ px (after remapping) and convolved with a Gaussian with $2.4$~Mm FWHM to match the MDI sampling and PSF.}
\label{fig_5}
    \end{figure*}

The latitudinal dependencies of $\langle \tau^\textrm{oi}\tau^\circlearrowleft\rangle$ and $\langle \tau^\textrm{oi}\tau^\circlearrowright\rangle$ for HMI data are plotted in Fig.~\ref{fig_5}a. Note that we filtered the travel-time maps spatially by removing the power for $kR_\odot > 300 $ since there is no significant signal in $\tau^\text{ac}$ at high wavenumbers. $\langle \tau^\textrm{oi}\tau^\circlearrowleft\rangle$ is negative in the northern hemisphere and positive in the southern hemisphere. For $\langle \tau^\textrm{oi}\tau^\circlearrowright\rangle$ the pattern is reversed. At the equator, both quantities have a low positive value of $2$~s$^2$, which is probably due to wave-speed perturbations associated with the magnetic network.

In Fig.~\ref{fig_5}b, $\langle \tau^\textrm{oi}\tau^\text{ac}\rangle$ is plotted versus solar latitude for both HMI and MDI data. This allows to compare the sensitivity of both instruments. Additionally, $\langle \tau^\textrm{oi}\tau^\text{ac}\rangle$ was computed from HMI Dopplergrams that were binned over $4\times 4$ pixels and convolved with a Gaussian with $2.4$~Mm FWHM (black curve) such that the spatial resolution matches approximately the resolution of MDI full-disk data. For both instruments, $\langle \tau^\textrm{oi}\tau^\text{ac}\rangle$ is negative in the northern hemisphere and positive in the southern hemisphere. As expected, $\langle \tau^\textrm{oi}\tau^\text{ac}\rangle$ is consistent with zero at the equator (no Coriolis force acts on horizontal flows). Our measurements are qualitatively consistent with the $\langle \text{div}_h \omega_z \rangle$ estimates by \citet{duvall_2000} and \citet{gizon_2003}.

Away from the equator, we see that the amplitude of $\langle \tau^\textrm{oi}\tau^\text{ac}\rangle$ is higher for HMI than for MDI by a factor of five. This means HMI has a much higher sensitivity to $\langle \tau^\textrm{oi}\tau^\text{ac}\rangle$ than MDI. Only when the HMI Dopplergrams are degraded to match the MDI sampling and PSF, a good agreement of the curves is achieved. Hence the difference in sensitivity to vertical vorticity between HMI and MDI is directly related to the spatial resolution of the instruments.

\begin{figure}
  \centering
\includegraphics[width=\hsize]{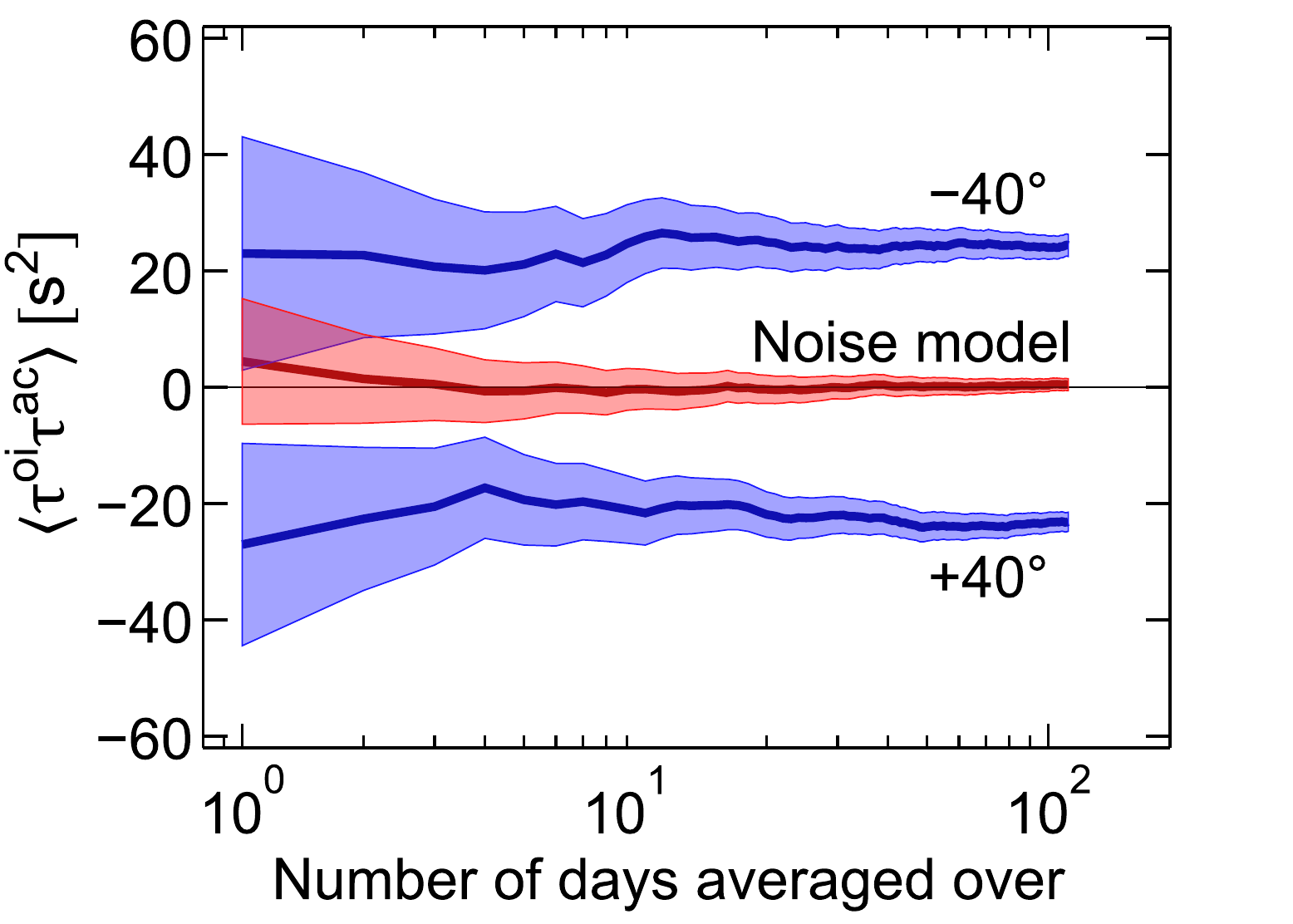}
   \caption{Product $\langle \tau^\textrm{oi}\tau^\text{ac}\rangle$ as a function of days of data averaged over, for f-mode-filtered HMI observations at $-40\degr$ and $40\degr$ latitude as well as $\tau^\text{ac}$ computed for the noise model. The filled areas denote the $3\sigma$ estimates. All data have been averaged over space (about $180 \times 180$~Mm$^2$), four angles $\beta$ and three 8~h datasets per day.}
\label{fig_6}
    \end{figure}

As in Sect.~\ref{chap_test1}, we investigate how much data is needed for a clear non-zero detection ($3\sigma$ level) of $\langle \tau^\textrm{oi}\tau^\text{ac}\rangle$ away from the equator. Figure~\ref{fig_6} shows $\langle \tau^\textrm{oi}\tau^\text{ac}\rangle$ versus the duration of the HMI observations at solar latitudes $\pm40\degr$. Again the $3\sigma$ error estimates are plotted as shaded areas. The curves for observations at $40\degr$ and $-40\degr$ latitude do not overlap after averaging over one or more days of data. This means a difference in $\langle \tau^\textrm{oi}\tau^\text{ac}\rangle$ between the hemispheres can be detected at the $3\sigma$ level after one day of averaging. To distinguish between $40\degr$ observations and the noise-model data takes roughly three days of data. Note that these results are only valid for the given map size of about $180 \times 180$~Mm$^2$. This corresponds to averaging over roughly 40 supergranules per day.

\section{Conclusion}
We have presented a new averaging scheme for time-distance helioseismology, which has direct sensitivity to the vertical component of the flow vorticity.  The anti-clockwise minus clockwise HMI travel-time maps for f modes show power above the noise level. Unlike the divergence signal, the vorticity signal does not peak at supergranular scales but increases continuously toward larger spatial scales. Furthermore, the latitudinal dependence of the correlation between the vorticity and the divergence signals is consistent with the effect of the Coriolis force on turbulent convection. We find that HMI has a much higher sensitivity to this correlation than MDI.


\begin{acknowledgements}
We acknowledge research funding by Deutsche Forschungsgemeinschaft (DFG) under grant SFB 963/1 ``Astrophysical flow instabilities and turbulence'' (Project A1). The HMI data used are courtesy of NASA/SDO and the HMI science team. The data were processed at the German Data Center for SDO (GDC-SDO), funded by the German Aerospace Center (DLR). We are grateful to C.~Lindsey, H.~Schunker and R.~Burston for providing help with tracking and mapping. We acknowledge the workflow management system Pegasus (funded by The National Science Foundation under OCI SI2-SSI program grant \#1148515 and the OCI SDCI program grant \#0722019).
\end{acknowledgements}

\bibliographystyle{aa}
\bibliography{literature}

\end{document}